\documentclass[amsmath,amssymb,showpacs,twocolumn,superscriptaddress,prl]{revtex4}

\usepackage{amsmath,amsthm,amssymb,bm,hyperref}

\newcommand{\be}{\begin{equation}}
\newcommand{\ee}{\end{equation}}
\newcommand{\fa}{\Psi}
\newcommand{\fad}{\Psi^\dagger}
\newcommand{\lam}{\lambda}
\newcommand{\ra}{\rangle}
\newcommand{\la}{\langle}
\newcommand{\inti}{\int_{-\infty}^{+\infty}}

\begin{document}

\title{Non-conformal asymptotic behavior of the time-dependent field-field
correlators of 1D anyons}

\author{Ovidiu I. P\^{a}\c{t}u}

\affiliation{C.N. Yang Institute for Theoretical Physics, State
University of New York at Stony Brook, Stony Brook, NY 11794-3840,
USA }
\affiliation{Institute for Space Sciences, Bucharest-M\u{a}gurele, R
077125, Romania}

\author{Vladimir E. Korepin}

\affiliation{C.N. Yang Institute  for Theoretical Physics, State
University of New York at Stony Brook, Stony Brook, NY 11794-3840,
USA }

\author{Dmitri V. Averin}
\affiliation{Department of Physics and Astronomy, State University
of New York at Stony Brook, Stony Brook, NY 11794-3800, USA }
%\email[Electronic addresses: ]{ipatu@grad.physics.sunysb.edu ;
%korepin@max2.physics.sunysb.edu ; dmitri.averin@stonybrook.edu}

\begin{abstract}
The exact large time and distance behavior of the field-field
correlators has been computed for one-dimensional impenetrable anyons
at finite temperatures. The result reproduces known asymptotics for
impenetrable bosons and free fermions in the appropriate limits of
the statistics parameter. The obtained asymptotic behavior of the
correlators is dominated by the singularity in the spectral density
of the quasiparticle states at the bottom of the band, and differs
from the predictions of the conformal field theory. One can argue,
however, that the anyonic response to the low-energy probes is still
determined by the conformal terms in the asymptotic expansion.
\end{abstract}


\maketitle

%%%%%%%%%%%%%%%%%%%%%%%%%%%%%%%%%%%%%%%%%%%%%%%%%%%%%%%%%%%%%%%%%%%%%

When confined to move in one dimension, hard-core impenetrable
particles cannot be exchanged directly. This mean that the symmetry
of the wavefunction with respect to exchanges of coordinates can be
defined arbitrarily, e.g., independently of the dynamic
particle-particle interaction (see, e.g., \cite{suth,LM}). The
wavefunction describes then the exchange statistics with, in general,
fractional statistics parameter $\kappa \in [0,1]$ interpolating
between Bose ($\kappa=0$) and Fermi ($\kappa=1$) particles.
Quantitatively, this ``anyonic'' exchange statistics implies that the
multi-particle wavefunctions behave as \cite{AK,MDG,AN}:
$\psi(\cdots,x_{i+1},x_i,\cdots)=e^{i\pi\kappa \epsilon
(x_i-x_{i+1})}\psi(\cdots,x_i,x_{i+1},\cdots)$, upon exchange of the
two nearest-neighbor coordinates, where $\epsilon(x)\equiv x/|x| ,\,
\epsilon(0)=0$. While the wavefunctions $\psi$ depend on the
statistics parameter $\kappa$, the absence of real exchanges means
that thermodynamic properties of the hard-core particles are
independent of statistics. In particular, in the simplest case of
only hard-core interaction of zero radius, which is considered in
this work, the thermodynamics coincides \cite{BGO,PKA} with that of
free fermions.

The exchange statistics manifests itself, however, in the correlation
functions of the field operators, which describe processes with
changing number of particles. In the general case of 1D anyons, only
``static'' same-time correlators of the field operators have been
calculated \cite{SSC,SC1,SC2,PKA3,PKA4} so far. The purpose of this
work is to present the exact calculation of the large time and distance
asymptotics of the time-dependent field-field correlators of
impenetrable anyons. The time-dependent correlators are more relevant
experimentally, as they determine, e.g., the tunneling density of
states of the system. For anyons, tunneling characteristics can be
measured, e.g., in FQHE structures (see, e.g., \cite{AN,CZG}). The
new qualitative feature of the time-dependent anyonic correlators
obtained in this work is that, in contrast to previously found
\cite{IIKV,KBI} asymptotics for impenetrable bosons, the correlators
for the general statistics parameter $\kappa$ do not exhibit
conformal behavior in the low-temperature limit. This result
challenges the accepted notion that the large-time asymptotic
behavior of the correlators of the 1D systems is determined by the
low-energy excitations. Our exact calculation shows that the
large-time asymptotic behavior can actually be dominated by the
singularity of the density of the quasiparticle states at the bottom
of the energy band which is characteristic for one dimension. This
conclusion adds another limitation on the applicability of the
conformal field theory to the large space and time asymptotic
description of the 1D particle systems. Previously known limitations
were related to the effects of the strong interaction on particles
with spin, which leads to vanishing of the exchange energy scale,
making the spin dynamics non-conformal \cite{AB,ANZ,CZ,GAF}, and the
effects of finite nonlinearity of the quasiparticle spectrum at the
Fermi energy \cite{KPKG}.

In details, {\em the model} considered in this work describes
one-dimensional hard-core anyons interacting via a $\delta$-function
potential, which simulates the hard-core repulsion, and are free
otherwise. The second quantized Hamiltonian of such anyons is given
by
\begin{eqnarray}
H&=&\int dx \left [\partial_x \fad(x)\partial_x\fa(x)
+c\fad(x)\fad(x)\fa(x)\fa(x) \right. \nonumber \\ & &\ \
\;\;\;\;\;\;\; \left. -h\, \fad(x)\fa(x) \right] , \label{hama}
\end{eqnarray}
in the $c\rightarrow \infty$ limit. Here $h$ is the chemical
potential, and the anyonic fields satisfy the following commutation
relations
\begin{eqnarray}
\fa(x_1)\fad(x_2)&=&e^{-i\pi\kappa\epsilon(x_1-x_2)} \fad(x_2)
\fa(x_1)+ \delta(x_1-x_2),\nonumber\\
\fa(x_1)\fa(x_2)&=&e^{i\pi\kappa\epsilon(x_1-x_2)} \fa(x_2)
\fa(x_1)\, . \label{com1} \end{eqnarray}
The focus of this work is on the calculation of the large time and
distance asymptotic behavior of the field-field correlation function
at finite temperature defined as
\[ \la\fa(x_2,t_2)\fad(x_1,t_1)\ra=\frac{\mbox{ Tr }(e^{-H/T}
\fa(x_2,t_2)\fad(x_1,t_1))}{\mbox{ Tr } e^{-H/T}}\, . \]
As a function of the statistics parameter $\kappa$, this correlator
should interpolate between the case of impenetrable bosons
($\kappa=0$) and free fermions ($\kappa=1$).

To state first our {\em main result}, we introduce the rescaled
variables $x=(x_1-x_2)\sqrt{T}/2>0,\ t=(t_2-t_1)T/2>0,\ \beta=h/T$.
In terms of these variables, the field correlator can be written as
\be \label{re} \la\fa(x_2,t_2)\fad(x_1,t_1)\ra=\sqrt{T}
g(x,t,\beta,\kappa)\, , \ee
where the function $g$ is defined below. Due to the fact that
$g(x,t,\beta,\kappa)=g(-x,t,\beta,-\kappa)$ and
$g(x,t,\beta,\kappa)=g^*(x,-t,\beta,-\kappa)$ it is sufficient to
consider only the case $x>0,\ t>0$. One needs to distinguish the
space-like, $x/2t>\sqrt{\beta}$, and time-like, $x/2t<\sqrt{\beta}$,
regions. While the leading term in the asymptotics is the same in
both regions, the next-to-leading term behaves in them differently.
The large time and distance asymptotic form of the function $g$, for
$x,t\rightarrow \infty $, with $x/t=const$, is:
\be \label{mainres} g(x,t,\beta,\kappa)\simeq
t^{\nu^2/2}e^{C(x,t,\beta,\kappa)+ix I(\beta,\kappa)} \ee

\vspace*{-4ex}

\[ \times [c_0 t^{-1/2-i\nu}e^{2it(\lam_s^2+\beta)}+
c_1 e^{2(\pm t\pi\kappa-ix\lam_0^{\mp})} +o(1/\sqrt{t} )] \, , \]
where
\[ \lam_0^{\mp}=[-\alpha^{1/2} \mp i\left(\alpha-2\beta
\right)^{1/2}]/\sqrt{2}\, , \;\;\; \alpha=\beta+ \sqrt{\beta^2
+\pi^2\kappa^2} , \]
and the upper (lower) sign corresponds to the space-like (time-like)
regions. Also, $\lam_s=-x/2t$, $\nu=-\frac{1}{\pi} \ln|\varphi(
\lam_s^2,\beta, \kappa)|$, the constants $c_0,c_1$ are some
undetermined amplitudes, $I(\beta,\kappa)=\Im \inti \ \ln
\varphi(\lam ^2,\beta,\kappa)\ d\lam /\pi $, and
\be C(x,t,\beta,\kappa)=\frac{1}{\pi}\inti|x-2t\lam|\ln| \varphi(
\lam^2,\beta,\kappa)|\ d\lam\, , \label{e4} \ee
with
\be \varphi(\lam^2,\beta,\kappa) \equiv (e^{\lam^2-\beta}-e^{i\pi
\kappa}) /(e^{\lam^2-\beta}+1)\, . \label{e5} \ee
Even though it seems that the second term (with amplitude $c_1$) in
the expansion (\ref{mainres}) is superfluous, because it is in
general exponentially decreasing and smaller than the error term, one
can see that it gives the dominant contribtion in the bosonic limit,
thus justifying its presence. Another remark is that the precise
analytic expression for the second term in the expansion
(\ref{mainres}) depends on the analytical structure of the solution
of the Riemann-Hilbert problem. While there is no additional
difficulties in analyzing all regimes, Eq.~(\ref{mainres}) is written
out specifically in the situation when this term is determined by the
pole at $\lam_0^{\mp}$, which is the case when
\be \label{cond} | \Re \sqrt{\beta+i \pi\kappa} -x/2t| >\Im
\sqrt{\beta+i \pi\kappa} \, . \ee
The positive branch of the square root should be taken in this
relation. This condition is always satisfied, in particular, in the
more interesting low-temperature regime, $\beta >>1$.

%%%%%%%%%%%%%%%%%%%%%%%%%%%%%%%%%%%%%%%%%%%%%%%%%%%%%%%%%%%%%%%%%%%
We begin the analysis of Eq.~(\ref{mainres}) by demonstrating that it
reproduces the known asymptotics for bosons and fermions. In the
{\em bosonic limit} $\kappa\rightarrow 0$, we have $\varphi(\lam^2,
\beta, 0)=(e^{\lam^2-\beta}-1)(e^{\lam^2-\beta}+1)$, $\nu=-(1/\pi)
\ln| \varphi(\lam_s^2,\beta,0)|$, and
\[ C(x,t,\beta,0)=\frac{1}{\pi}\inti|x-2t\lam|\ln \left| \frac{
e^{\lam^2-\beta}-1}{e^{\lam^2-\beta}+1}\right|\, d\lam . \]
Also, $I(\beta,0)=-2 \sqrt{\beta}$ for $\beta>0$, and $I(\beta,0)=0$
for $\beta<0.$ In the case of negative chemical potential,
$\beta<0$, condition (\ref{cond}) implies that the second term in
the expansion (\ref{mainres}) is exponentially small and the
asymptotic behavior is determined by the first term, which reduces
for $\kappa=0$ to:
\[ g(x,t,\beta)\simeq c_0t^{(\nu-i)^2/2}e^{C(x,t,\beta,0)
+2it(\lam_s^2+\beta)} . \]
For the positive chemical potential, however, $\lam_0^{\mp}=
-\sqrt{\beta}$, and the leading contribution in both space-like and
time-like regions is given by the second term in the expansion
(\ref{mainres}), thus giving for $\beta=h/T>0$:
\[ g(x,t,\beta)\simeq c_1 t^{\nu^2/2}e^{C(x,t,\beta,0)}.
\]
Both of these expressions agree with the previous results
\cite{IIKV,KBI} for the impenetrable bosons.

In the case of  {\em free fermions}, the equivalent $g$ function in
the rescaled variables is
\[ g(x,t,\beta)=\frac{e^{2it\beta}}{2\pi}\inti \frac{e^{\lam^2
-\beta}}{e^{\lam^2-\beta}+1}e^{-2it\lam^2-2ix\lam}\ d\lam\, . \]
In the large $t$ and $x$ limit with $t/x=const$, the steepest
descent method gives the following asymptotic behavior under the
condition (\ref{cond}):
\[ g(x,t,\beta)\simeq c_0 t^{-1/2}e^{2it(\beta+\lam_s)}+c_1
e^{2(\pm t\pi-ix\lam_{f}^{\mp})}  . \]
Here $\lam_f^{\mp}$ are given by the same expression as
$\lam_0^{\mp}$ with $\kappa=1$, and as before, the upper (lower)
sign corresponds to the space-like (time-like) region. Comparing
this expression to Eq.~(\ref{mainres}), which is simplified for
$\kappa\rightarrow 1$ by the relations $C(x,t,\beta,1)= I(\beta,1)
=\nu=0$, we see that Eq.~(\ref{mainres}) indeed reproduces the
correct behavior of the asymptotic correlators for free fermions.

At positive chemical potential and low temperatures the system is
expected to exhibit {\em conformal behavior}. One of the simplest
ways to calculate the asymptotic correlators in this regime is to
use the standard bosonization approach \cite{FDMH}, in which the
field operators are expressed through two bosonic fields: the
integral $\theta(x)$ of the long-wave part of the density
fluctuations $\rho (x)$, i.e., $\partial \theta(x)/\partial x=\rho
(x)/\pi $, and the conjugate field $\phi (x)$ defined by the
commutation relation $[\phi (x), \theta (x')]=i\pi
\epsilon(x-x')/2$. The fact that the operator $e^{i\phi}$ reduces
density by the amount that corresponds to one particle, and $\theta$
produces the appropriate phase changes of the wavefunction across
each particle, implies then that the anyonic operators (\ref{com1})
can be written as \cite{CM}
\be \fa (x,t) \sim \sum_{m}e^{-i(\kappa+2m) [k_Fx+\theta(x,t)]} \,
e^{i \phi (x,t)} \, . \label{bos1} \ee
The sign $\sim$ in Eq.~(\ref{bos1}) is the reminder that the
relative amplitudes of different components are not defined in the
sum. The commutation relations between $\theta$ and $\phi$ shows
directly that the individual terms in (\ref{bos1}) satisfy the
appropriate exchange relations (\ref{com1}) away from $x_1=x_2$. The
standard average over the equilibrium fluctuations of $\theta$ and
$\phi$ gives then the large time and distance behavior of the
correlator of the anyonic field operators (\ref{bos1})
\cite{CM,PKA}. The two leading asymptotic terms are:
\[ \la \fa(x_2,t_2)\fad(x_1,t_1) \ra = \frac{b_0 \, e^{i\kappa
k_Fx_{12}}}{[u_-]^{(1-\kappa)^2/4} [u_+]^{(1+\kappa)^2/4}}
\]

\vspace{-3ex}

\be +\, \frac{b_{-1} \, e^{i(\kappa-2) k_Fx_{12}}}{[u_-]^{(3- \kappa
)^2/4} [u_+]^{(1-\kappa)^2/4}} \, , \label{bos2} \ee
where $u_{\pm} \equiv \sinh[\pi T (t_{21}-i0 \pm x_{12}/v_F)]$, the
constants $b$ are some undetermined amplitudes of different
components of the correlator, and $t_{21}= t_2-t_1$, $x_{12}=
x_1-x_2$. The Fermi vector and the Fermi velocity are, respectively,
$k_F=\sqrt{h}$ and $v_F=2\sqrt{h}$.

In Equation (\ref{bos2}), we kept the two leading terms to facilitate
comparison to the case of the static correlators \cite{PKA4}, for
which both terms are important in the vicinity of the Fermi point
$\kappa=1$. For the dynamic correlators considered in this work, it
is assumed that $t_{21} \neq 0$. In this case, only the first term is
relevant for our discussion. In the $T\rightarrow 0$ limit, the
criterion used above to separate the space-time and time-like regions
reduces to the comparison between $t_{21}$ and $x_{12}/v_F$. Keeping
only the leading term in Eq.~(\ref{bos2}) we see that this equation
reduces to
\[ \la\fa(x_2,t_2)\fad(x_1,t_1)\ra \simeq e^{ik_F\kappa x_{12}}
e^{\pi T\kappa t_{21}} e^{-\frac{\pi T } {v_F}x_{12}\left(
\frac{1}{2}+\frac{\kappa^2}{2}\right)}, \]
in the space-like region, and to
\[ \la\fa(x_2,t_2)\fad(x_1,t_1)\ra \simeq e^{ik_F\kappa
x_{12}}e^{\frac{\pi T\kappa}{v_F}x_{12}} e^{-\pi T t_{21}
\left(\frac{1}{2}+\frac{\kappa^2}{2}\right)} , \]
in the time-like region. In the same low-temperature limit, one can
also simplify our main expression (\ref{mainres}). Indeed, for
$T\rightarrow 0$ one has: $ix I(\beta,\kappa) \simeq
2ix\sqrt{\beta}(\kappa-1)$, and $\lam_0^{\mp} \simeq
-\sqrt{\beta}\mp i\frac{\pi\kappa}{2\sqrt{\beta}}$. Also,
\[ C(x,t,\beta,\kappa)=- \pi (1-\kappa)^2 \left\{ \begin{array}{cc}
x/(2\sqrt{\beta}) , & x/2t> \sqrt{\beta}\, , \\
t , &  x/2t<\sqrt{\beta} \, . \end{array}    \right. \]
Using these equations one can see directly that the exponential parts
of the next-to-leading term in Eq.~(\ref{mainres}) combine to give
exactly the same result as predicted by CFT. This means that although
the leading asymptotic term in the correlator (\ref{mainres}) is not
conformal, the next-to-leading term is. The exponential factors in
the leading term of the correlator (\ref{mainres}) imply that in the
low temperature limit, $\beta \gg 1$, this term can manifest itself
in the response of the anyonic liquid to external perturbation only
in the case of large energy $\omega$ of the perturbation $\omega
\simeq \epsilon_F=h$. Therefore, at small energies, $\omega \ll
\epsilon_F$, the response is determined by the second asymptotic
term, and coincides with the prediction of the CFT.

To see this more explicitly, one can consider the simplest example of
weak single-point tunneling into the anyonic liquid. As usual, the
tunneling density of states is given in this case by the Fourier
transform of the field correlator with $x=0$. Equation
(\ref{mainres}) for the anyonic correlator reduces for $x=0$ (as
before, $\beta \gg 1$) to:
\be \label{tunn} g(0,t,\beta,\kappa) \simeq e^{C(0,t,\beta,\kappa)}
[c_0 (e^{2it \beta}/\sqrt{t}) + c_1 e^{-2 t\pi\kappa } ]\, . \ee
Taking Fourier transform of this correlator, one can see that the
contribution of the first term to the tunneling density of states
$A(\omega)$ is $A(\omega) \propto (h-\omega)^{-1/2}$. Although the
large-time asymptotics (\ref{tunn}) is not sufficient to establish
all features of $A(\omega)$, this estimate illustrates the fact that
this term in the asymptotic correlator is produced by the
singularity of the 1D density of states at the bottom of the band.
This implies that at small energies this contribution to tunneling
will be suppressed, and the tunneling density of states will exhibit
the power-law nonlinearities with the power $(1+\kappa^2)/2$
produced by the leading term in the CFT correlator (\ref{bos2}).

As the last part of the discussion of our main result
(\ref{mainres}), we provide a simple heuristic interpretation of the
exponential factors in the leading term in this equation. The qualitative
statement proven by this interpretation is that in the $T\neq 0$
regime considered in this work, the exponential decay of the
correlator is caused mainly by the thermal fluctuations of the number
of particles which create the fluctuations of the accumulated
statistical phase. The average taken over these phase fluctuations
leads to the exponential suppression of the correlator. To see this,
one starts with the standard Wigner-Jordan transformation expressing
the anyonic fields (\ref{com1}) in terms of (in our case free)
fermions $\xi$:
\[ \fa(x,t)=e^{i \pi (1-\kappa)n(x,t)}\xi(x,t), \;\;\; n(x,t)
=\int^{x} dx'\rho(x',t)  \, , \]
where $\rho=\xi^{\dagger}\xi$ is the operator of the particle
density. This transformation results in the following expression for
the anyonic field-field correlator:
\[ \la \fa(x_2,t_2)\fad(x_1,t_1)\ra= \la \xi(x_2,t_2) e^{i \pi
(\kappa-1)n_d} \xi^{\dagger}(x_1,t_1) \ra , \]
where $n_d\equiv n(x_1,t_1)-n(x_2,t_2)$. For large distances
$x_1-x_2$ and time differences $t_2-t_1$, the fluctuations of $n_d$
can be calculated from the following quasiclassical considerations.
Contribution to these fluctuation from particles with different
momenta $k$ can be treated independently from each other. For given
$k$, the particle fluctuations at the initial moment $t_1$ are
governed by the Fermi distribution $\vartheta(k)= 1/[e^{(k^2
-h)/T}+1]$, which gives the probability density
$\vartheta(k)dk/2\pi$ for momentum-$k$ particles to be present in
any space interval. The resulting random distribution of particles
moves with time with velocity (in our notations) $2k$. The
difference $n_d$ can then take values 1,0, or -1, producing the
phase factors in the correlator $-e^{i \pi \kappa }, 1, -e^{-i \pi
\kappa }$. Analyzing the shifts of the distributions with time, and
averaging over the probabilities to have or not to have a particle
in the initial distribution, one sees that
\be  \la e^{i \pi (\kappa-1) n_d } \ra = e^{C(x,t,\beta,\kappa)+ix
I(\beta,\kappa)} ,  \label{fer3} \ee
demonstrating that this part of the asymptotic suppression of the
anyonic correlator is indeed caused by thermal fluctuations of the
statistical phase.

Qualitatively, the non-conformal contribution to the asymptotic
behavior of the anyonic field-field correlator found in this work is
analogous to the power-law corrections to the exponential decay of
the metastable states in general quantum mechanical problems (see,
e.g., \cite{LMK} and references therein). Similarly to these general
corrections \cite{KU}, which also result from the bounded-from-below
nature of the energy spectrum of a quantum system, our non-conformal
power-law terms in the asymptotics come with the energy shift to the
bottom of the band. This shift ensures that despite the presence of
the non-conformal terms, the anyonic response at energies close to
the Fermi energy, agrees, as demonstrated explicitly by our results,
with the predictions of the conformal field theory.

%%%%%%%%%%%%%%%%%%%%%%%%%%%%%%%%%%%%%%%%%%%%%%%%%%%%%%%%%%%%%%%%%%%
Finally, we present a brief {\em outline of the derivation} of the
asymptotic behavior of the correlator (\ref{mainres}). The starting
point of our calculations is the determinant representation of the
space-, time-, and temperature dependent field-field correlator for
anyons obtained in \cite{PKA3}. In rescaled variables, the
correlator is given by Eq.~(\ref{re}), with
\be\label{defg}
g(x,t,\beta,\kappa)=-\frac{1}{2\pi}e^{2it\beta}b_{++}\det(1+\hat V_T)\, .
\ee
In Equation (\ref{defg}), $\hat V_T$ is an integral operator acting
on the entire real axis with the kernel given by
\[ V_T(\lam,\mu)=\frac{e_+(\lam)e_-(\mu)-e_-(\lam)e_+(\mu)}{\lam-
\mu}\, , \]
where
\[ e_-(\lam)=\frac{\cos(\pi\kappa/2)}{\pi}\sqrt{\vartheta(\lam)}
e^{i\phi(\lam)}\, ,\ \ e_+(\lam)=e_-(\lam)E(\lam)\, , \]
\[ E(\lam)=\mbox{P.V.} \int_{-\infty}^\infty d\mu\
 \frac{e^{-2i\phi(\mu)}}{\mu-\lam}+
 \pi\tan( \pi \kappa/2)e^{-2i\phi(\lam)}\, , \]
and $\phi(\lam)=t\lam^2+x\lam$. The auxiliary potential $b_{++}$  is
defined as $b{++}=B_{++}-G$ with $G=\inti e^{-2i\phi(\lam)}\ d\lam$
and $B_{++}=\inti e_{+}(\lam)(1+\hat V_T)^{-1} e_{+}(\lam)\ d\lam $.
The integral operator appearing in Eq.~(\ref{defg}) is of the special
kind  called ``integrable" operators\cite{KBI,HI}. This type of
operators appears frequently in the investigations of correlation
functions of integrable models and distribution of eigenvalues of
random matrices \cite{HI}. The specific factorization properties of
the kernel $V_T$ allow one to obtain differential equations for the
correlators. These differential equations do not depend on the
statistics parameter $\kappa$ and are the same  as the ones obtained
for impenetrable bosons. The statistics parameter enters in the
initial conditions \cite{PKA4,SC1}. The large time and distance
asymptotic behavior of the correlation functions is obtained from the
analysis of a matrix Riemann-Hilbert problem associated with these
differential equations using the ``Manakov ansatz" \cite{M,I,
IIKV,KBI}. The details of the computations will be presented
elsewhere.

The authors were supported by the NSF grants PHY-0653342,
DMS-0503712 and DMR-0325551 .

\end{document}